\documentclass[aps,prd,superscriptaddress,twocolumn]{revtex4-1}

\usepackage{graphicx}
\usepackage{amsmath}
\usepackage{amssymb}
\usepackage{color}
\usepackage[per-mode = fraction]{siunitx}
\usepackage{bm}
\usepackage{placeins}
\usepackage{ulem}

\date{\today}

\begin{document}

\title{Work function, deformation potential, and collapse of Landau levels in strained graphene and silicene}

\author{D. Grassano}
\affiliation{Dept. of Physics, and INFN, University of Rome Tor Vergata, Via della Ricerca Scientifica 1, I-00133 Rome, Italy}
\email{davide.grassano@roma2.infn.it}

\author{M. D'Alessandro}
\affiliation{Istituto di Struttura della Materia-CNR (ISM-CNR), Via del Fosso del Cavaliere 100, 00133 Rome, Italy}

\author{O. Pulci}
\affiliation{Dept. of Physics, and INFN, University of Rome Tor Vergata, Via della Ricerca Scientifica 1, I-00133 Rome, Italy}

\author{S.G. Sharapov}
\affiliation{Bogolyubov Institute for Theoretical Physics, National Academy of Science of Ukraine, 14-b
Metrolohichna Street, Kyiv 03143, Ukraine}

\author{V.P. Gusynin}
\affiliation{Bogolyubov Institute for Theoretical Physics, National Academy of Science of Ukraine, 14-b
Metrolohichna Street, Kyiv 03143, Ukraine}

\author{A.A. Varlamov}
\affiliation{CNR-SPIN, c/o DICII-University of Rome ``Tor Vergata'', Via del Politecnico, 1, 00133 Rome, Italy}

\begin{abstract}
We  perform  a  systematic {\it ab  initio} study  of  the  work function  and  its uniform strain  dependence for
graphene and silicene for both tensile and compressive strains.
The Poisson ratios  associated with armchair and zigzag strains are also computed.
Based on these results, we obtain the  deformation potential, crucial for straintronics,  as a function of the applied strain.
Further, we propose a particular experimental setup with a special strain configuration that generates only the electric field,
while the pseudomagnetic field is absent.
Then, applying a real magnetic field, one should be able to realize experimentally  the spectacular phenomenon of the collapse
of Landau levels in graphene or related two-dimensional materials.
\end{abstract}

\maketitle

\section{Introduction}

One of the remarkable features of two-dimensional (2D) Dirac materials such as graphene and
silicene is the formation of the relativisticlike Landau levels in a magnetic field. When
in addition to the magnetic field
the  electric field is applied in the plane  the spectacular phenomenon  of Landau level collapse occurs.
It consists of the merging of the  Landau level staircase when the applied
electric field reaches  its  critical value and the cyclotron frequency becomes zero \cite{Lukose2007PRL,Peres2007JPCM}.
The behavior of Landau levels in these materials can be controlled by tuning  external magnetic and electric fields
along with strain that effectively induces artificial electromagnetic  fields  \cite{Vozmediano2010PR,Amorim2016PR,Naumis2017RPP}.

The value of the strain-induced electric field is determined by the corresponding {\it deformation potential}
that is present in the tight-binding description. In their turn the characteristics of this potential can be extracted
from the strain dependence of the {\it work function} (WF).
The relationship between the deformation potential and the WF can be determined accurately by both  experiments and
{\it ab initio} calculations.
Here we quantitatively evaluate the deformation potential by {\it ab initio}
calculation of the work function in the strained graphene and
silicene. Combining these systematic {\it ab initio} calculations of the WF,
the deformation potential, and the tight-binding model Hamiltonian, we propose
the new experimentally feasible condition of Landau level collapse in
graphene using strain.

{ (i)}
It is well known that the energies of relativistic Landau levels of graphene in a magnetic field $B$ applied perpendicular
to the plane in the presence of an in-plane electric field $E$ are
\cite{Lukose2007PRL,Peres2007JPCM}
\begin{equation}
\label{LL}
E_n =  \pm \Omega_L \sqrt{n} - \hbar k \frac{E}{B},
\end{equation}
where $k$ is the in-plane wave vector along the direction perpendicular to the electric field
and the Landau scale is
\begin{equation}
\label{Landau-scale}
\Omega_L =  \omega_L \left[1 - \frac{E^2}{v_F^2 B^2} \right]^{3/4}.
\end{equation}
Here $v_F $ is the Fermi velocity in graphene and
$\omega_L =\sqrt{2 |eB| \hbar v_F^2} $
is the Landau scale in the absence of an electric field.
The Landau level collapse occurs at the critical value $E_c = v_F B$ and the cyclotron frequency $\Omega_L/\hbar$ becomes zero.

{ (ii)} A new branch of study called {\it straintronics}   explores the possibilities to use strain for controlling the
physical properties of graphene and related materials  \cite{Vozmediano2010PR,Amorim2016PR,Naumis2017RPP}.
The electronic properties are probably the most desirable to control.
In particular, it allows one to govern the formation and behavior of the Landau levels
by means of  strain-induced  artificial magnetic and electric fields.

These strain effects have been investigated in the tight-binding
model.
It turns out that
the influence of deformation on the parameters of the model is mainly twofold.
First, the hopping integrals that describe the motion of conducting electrons between the atoms change under  strain.
For  uniform strain this results in a linear change of the slope of the  density of states function in the vicinity of the Dirac point.
Second, the on-site energies of the electrons (the deformation potential)
change, causing a shift of the Dirac point energy $E_\mathrm{D}^\varepsilon$ itself,
where  $\varepsilon$ is the strain.
As mentioned above, this potential can be extracted from the strain dependence of the WF.

{(iii)} The feasibility to tune  the WF of graphene and related new materials is important for engineering new efficient devices.
These  require a cathode and an  anode  electrodes with  low and high values of the WF, respectively.
The WF of the system $W = E_{\mathrm{vac}} -E_F$ is defined as the difference between the values of the vacuum level
$E_{\mathrm{vac}}$ and the Fermi energy $E_F$ \cite{Cahen2003AM}.
The experimental value of the WF extrapolated for pristine undoped graphene is
$W \sim \SI{4.5}{eV}$ \cite{Yu2009NanoLett,Yan2012APL,Xu2013NanoLett}, which turns out to be in between the desired cathode
and anode values.

The WF of single and double  graphene layers can be varied by electrostatic gating \cite{Yu2009NanoLett} that changes the doping level.
It has to be stressed that the tunability of the WF by electrostatic gating in 2D materials is a rather nontrivial property. Indeed,  it is well known that in  most of the three-dimensional semiconductors the phenomenon of surface state pinning of the Fermi level occurs.
Here any change in $E_F$ is accompanied by an almost equal shift in the band structure and thus in the value of $E_{\mathrm{vac}}$ at the surface.
On the contrary, as it was demonstrated in Ref.~\cite{Samaddar2016Nanoscale},  in monolayer graphene the WF varies in one-to-one correspondence with the position of the Fermi level with respect to the Dirac point $E_D$.
This relation was verified down to the nanometer scale where, due to inhomogeneities of the sample, the local Dirac point also changes its position.

It is demonstrated that the WF of chemically vapor-deposited graphene can be adjusted by applying  strain \cite{He2015APL},
viz., under a 7\% uniaxial strain it increases by $\SI{0.16}{eV}$.
Finally, the WF of suspended exfoliated graphene
\cite{Volodin2017APL} increases by $\SI{18}{meV}$ under a strain of 0.3\% while the {\it ab initio} calculations
\cite{Volodin2017APL} found a $1.4$ times stronger effect.

The deformation potential is  a key input parameter for designing novel nanodevices that exploit the
strain tunability of the WF. To illustrate the significance of this parameter in the present work, we consider a scenario,
where it determines the value of the critical magnetic field $B_c$ when the  Landau levels in graphene collapse.

In the present work we propose to create a strain configuration that generates the electric field only,
while  the  pseudomagnetic \cite{Vozmediano2010PR,Amorim2016PR} field  is absent.
The strength of the electric field is governed by the deformation potential.
Implementing this  particular strain configuration along with applying a real magnetic  field, one would be able to realize
the phenomenon of the Landau level collapse more easily,  since the strain and the magnetic field can be independently varied.
The collapse in this case can be more easily controlled by the fine-tuning of
the external magnetic field.
This proposal differs from the previous work
\cite{Castro2017PRB}, where it was suggested to generate both
the electric field and  the pseudomagnetic fields by applying strain.
The Landau level collapse could not be easily tuned in the latter case.

Thus the outline of this work is threefold, viz., to present
a systematic {\it ab initio} study of the WF and its strain dependence for graphene and silicene, and to
extract the deformation potential from the WF. Finally, we consider a strain configuration such that
the potential strength determines the condition for
the Landau level collapse.

The paper is organized as follows.
In Sec.~\ref{sec:methods} we discuss the methods employed to study the problem including the full
{\it ab initio} computation of the Poisson ratio.
The results of our {\it ab initio} calculations are presented in Sec.~\ref{sec:ab-results}.
The strain dependencies of the WFs for graphene and silicene are discussed in
Secs.~\ref{sec:graphene} and \ref{sec:silicene}, respectively. These results are then used in
Sec.~\ref{sec:deformation-potential} to discuss the deformation potential part of the tight-binding Hamiltonian.
In Sec.~\ref{sec:discussion} we propose how to realize the phenomenon
of the Landau level collapse.
Finally, in the Conclusions  (Sec.~\ref{sec:concl}), we summarize  the obtained results.

\section{{\it Ab initio} methods}
\label{sec:methods}

The vast majority of the existing  tight-binding calculations on  strained graphene focus on the change of the hopping integrals  while neglecting the shift of the Dirac point energy $E_\mathrm{D}^\varepsilon$ itself,
where  $\varepsilon$ is the strain.
This one-sidedness probably explains why there is still no agreement on the value of the deformation potential that characterizes the strain
dependence of the Dirac point energy.
In particular, this effect is not mentioned  in Ref.~\onlinecite{Naumis2017RPP}, while theoretical values of the deformation potential  $\alpha$ recited in  Ref.~\onlinecite{Amorim2016PR} are rather inconsistent between different sources and vary in a fairly wide range from \SIrange{0}{20}{eV}.
A first-principles method to evaluate the deformation potential in realistic
material is much needed.

We consider the case of undoped graphene. The strain-induced shift of the
Fermi energy with respect to the vacuum level is evaluated. The physical meaning of the  deformation potential can be immediately understood from the fact that for
the undoped graphene the Fermi level coincides with the Dirac point $E_F = E_\mathrm{D}^\varepsilon$.
Then its  WF is $W_\mathrm{D}^\varepsilon = E_{\mathrm{vac}} -E_\mathrm{D}^\varepsilon$ and the deformation potential $\alpha$ characterizes the slope of its dependence on the strain, viz., $\alpha = - d W_\mathrm{D}^\varepsilon/d \varepsilon$.
The WF and its strain dependence can be found from the {\it ab initio} studies as suggested in \cite{Guinea2010PRB}.
In particular, using the result of the {\it ab initio} calculations \cite{Choi2010PRB}, which show that a $12\%$ uniaxial strain results in an increase of the work function $W_\mathrm{D}^\varepsilon $ by $\SI{0.3}{eV}$, one can estimate that  $\alpha   \approx \SI{-2.5}{eV}$.
This is a rather large value that implies that the impact of the deformation potential cannot be neglected, since even a moderate  $1\%$ strain causes an observable shift of the Dirac point by $\SI{25}{meV}$.
Concerning the other related 2D materials,
it is found in a recent first-principles density functional theory study \cite{Lanzillo2015JPCM}
that compressive strain of up to $10\%$ decreases the WF of various metal dichalcogenide monolayers
by as much as $\SI{1}{eV}$.

Our {\it ab initio} calculations of the effect of strain on the work function of graphene and silicene are
based on the density functional theory  as implemented in the QUANTUM ESPRESSO package \cite{espresso.2009,espresso.2017}.
We  solve the single-particle Schr\"{o}dinger equation as formulated by Kohn-Sham (KS) \cite{kohn.sham.1965}
\begin{equation}
\label{eqn:Kohn-Sham}
\begin{split}
\! \! \! \! \! \!
\left(\! -\!\frac{\hbar^2}{2m}\nabla^2\! +\! v_{\mathrm{ext}}({\bf r})\! +\! \int \! \frac{n({\bf r'})}{\left| {\bf r} \!- \!{\bf r'} \right|} d{\bf r'} \!+\! v_{\mathrm{xc}}({\bf r})\!\right)\psi^{\mathrm{K\!S}}_{i,{\bf k}}\!=\!\varepsilon^{\mathrm{K\!S}}_{i,{\bf k}} \psi^{\mathrm{K\!S}}_{i,{\bf k}},
\end{split}
\end{equation}
where $v_{\mathrm{ext}}$ is the electron-ion potential and $v_{\mathrm{xc}}$ is the exchange-correlation (XC) potential.
The Kohn-Sham equations are solved self-consistently through the wave function expansion on plane-wave basis sets with the use
of the periodic boundary conditions.
We use a 12$\times$12$\times$1 $k$-point mesh and an energy cutoff of \SI{100}{Ry}.
Periodic images of the 2D systems are separated along $z$ by a \SI{15}{\AA} vacuum, that turns out to be
a sufficient spacing to avoid spurious interaction among the image layers.
For the exchange-correlation potential $v_{xc}$, two functionals are used, namely, the local density approximation (LDA) and the Perdew-Burke-Ernzerhof functional (PBE) \cite{LDA, Perdew.Burke.ea:1996}.

As previously stated, the WF is calculated as the energy difference between the vacuum level and the Fermi energy.
To obtain the vacuum level we  compute the plane-averaged electrostatic potential associated with the ground-state
density of the system. Then, the vacuum level is given by the limit value of the potential at a far distance from the material.

Biaxial and uniaxial strains are applied by modifying the relative position of the atoms in
the lattice. The coordinate system is chosen in such a way that the zigzag direction is parallel to the $y$
axis and the armchair one is parallel to the $x$ axis. A generic uniform deformation is represented by
the strain matrix $\hat\varepsilon$:
\begin{equation}\label{eq:strain1}
\hat{\varepsilon} =
\begin{pmatrix}
\varepsilon_{xx} & \varepsilon_{xy} \\
\varepsilon_{yx} & \varepsilon_{yy} \\
\end{pmatrix},
\end{equation}
In this way the deformation of the lattice is described as
\begin{equation}\label{eq:strain2}
\bf{x} = \bf{x}_0 + \bf{u}, \qquad
\bf{u} = \hat{\varepsilon} \cdot \bf{x}_0,
\end{equation}
where $\bf{x}_0$ is the actual position of the atom and $\bf{u}$ is the displacement vector.
Specifically, the strain matrices describing the deformations for both biaxial strain and uniaxial strains
in the armchair and zigzag directions are given by
\begin{equation}
\begin{split}
\label{eq:poisson_ratio1}
\hat{\varepsilon}_{\mathrm{bi}} & =
\eta
\begin{pmatrix}
 1 & 0 \\
 0 & 1
\end{pmatrix},\\
\hat{\varepsilon}_{\mathrm{arm}} =
\eta
\begin{pmatrix}
 1 & 0 \\
 0 & -\nu_{\mathrm{arm}}
\end{pmatrix}
&, \quad
\hat{\varepsilon}_{\mathrm{zig}} =
\eta
\begin{pmatrix}
-\nu_{\mathrm{zig}} & 0 \\
0 & 1
\end{pmatrix},
 \end{split}
\end{equation}
where the strain parameter $\eta$ determines the magnitude of deformation and $\nu_{\mathrm{arm}}$ and
$\nu_{\mathrm{zig}}$ represent the Poisson ratios (PRs) associated with armchair strain and zigzag strain, respectively.
These parameters indicate the amount of deformation in the transverse direction, with respect to the applied strain.

\begin{figure}[!h]
\raggedleft{
\includegraphics[width=1.\linewidth]{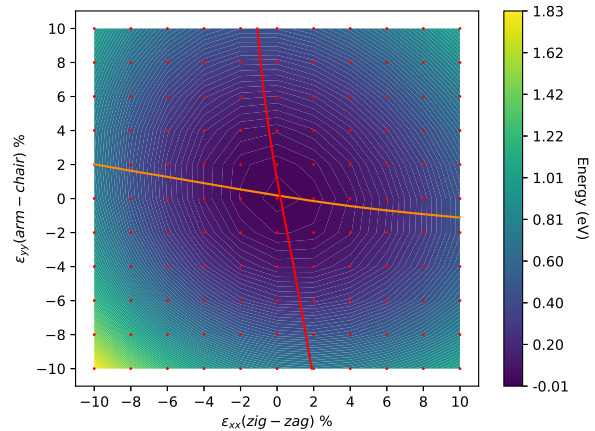}}
\caption{Plot of the total energy of graphene against the values of $\varepsilon _{xx/yy}$.
        The dark orange (red) curve represents the value of the $\varepsilon_{yy}$
        ($\varepsilon_{xx}$) strain that minimizes the energy of the system given the constraint
        $\varepsilon _{xx} (\varepsilon _{yy}) = \mbox{const}$.
        The red dots represent  the sampled combinations of $\epsilon_{xx/yy}$ values used to compute  the total energy of the system.
        }
\label{fig:graphene_poisson}
\end{figure}

In the present analysis the PR is computed in a full {\it ab initio} fashion.
We impose a fixed strain in a given direction and look for the value of the atomic distance in the
transverse coordinates that minimizes the total energy of the system. The results of this procedure are reported in Figs.~\ref{fig:graphene_poisson}  and \ref{fig:silicene_poisson} for graphene and silicene, respectively. This analysis provides values of the PR that depend on both the deformation direction and the strain value.
For silicene,  the corresponding buckling is determined for each applied strain.

\section{{\it Ab initio} results}
\label{sec:ab-results}

\subsection{Strain dependence of the WF and extraction of the deformation potential for graphene}
\label{sec:graphene}

\begin{figure}[!h]
\raggedleft{
\includegraphics[width=1.\linewidth]{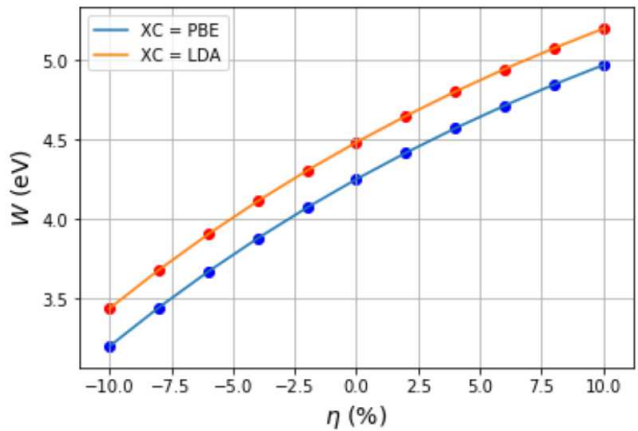}}
\caption{Dependence of the work function of graphene on the values of biaxial strain for the PBE and LDA XC functionals.}
\label{Wf_Xc_dependence}
\end{figure}

The strain dependence of the work function in graphene  is computed
using the PBE functional, since, as shown in Fig.~\ref{Wf_Xc_dependence} for biaxial strain,
the value of the WF turns out to be quite sensitive to the choice of the XC functional but its slope is not.
This implies that physical quantities like the deformation potential, which is basically related to the first
derivative of these curves, can be assessed with less ambiguity.

\begin{figure}[!h]
\raggedleft{
\includegraphics[width=1.\linewidth]{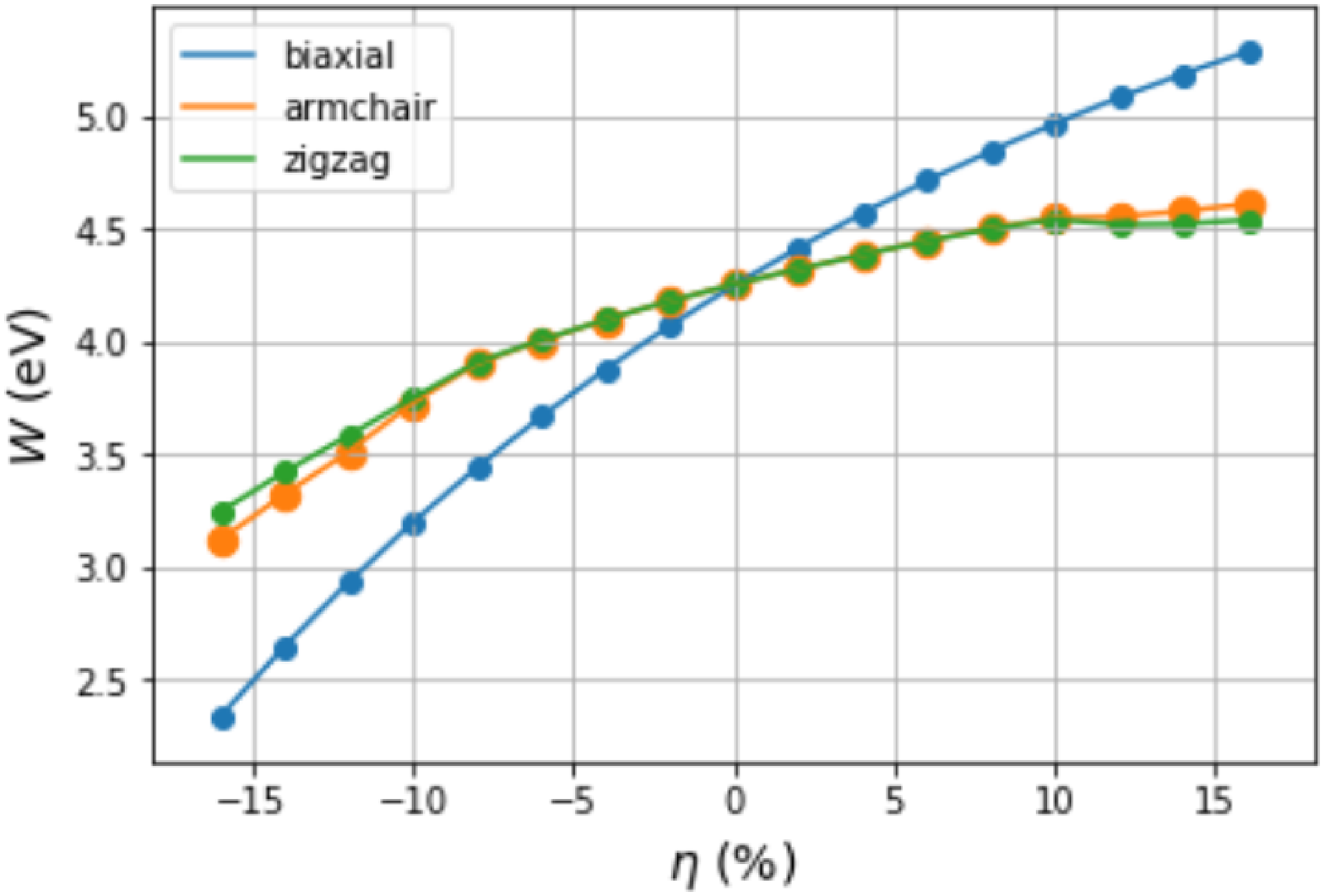}
\includegraphics[width=1.\linewidth]{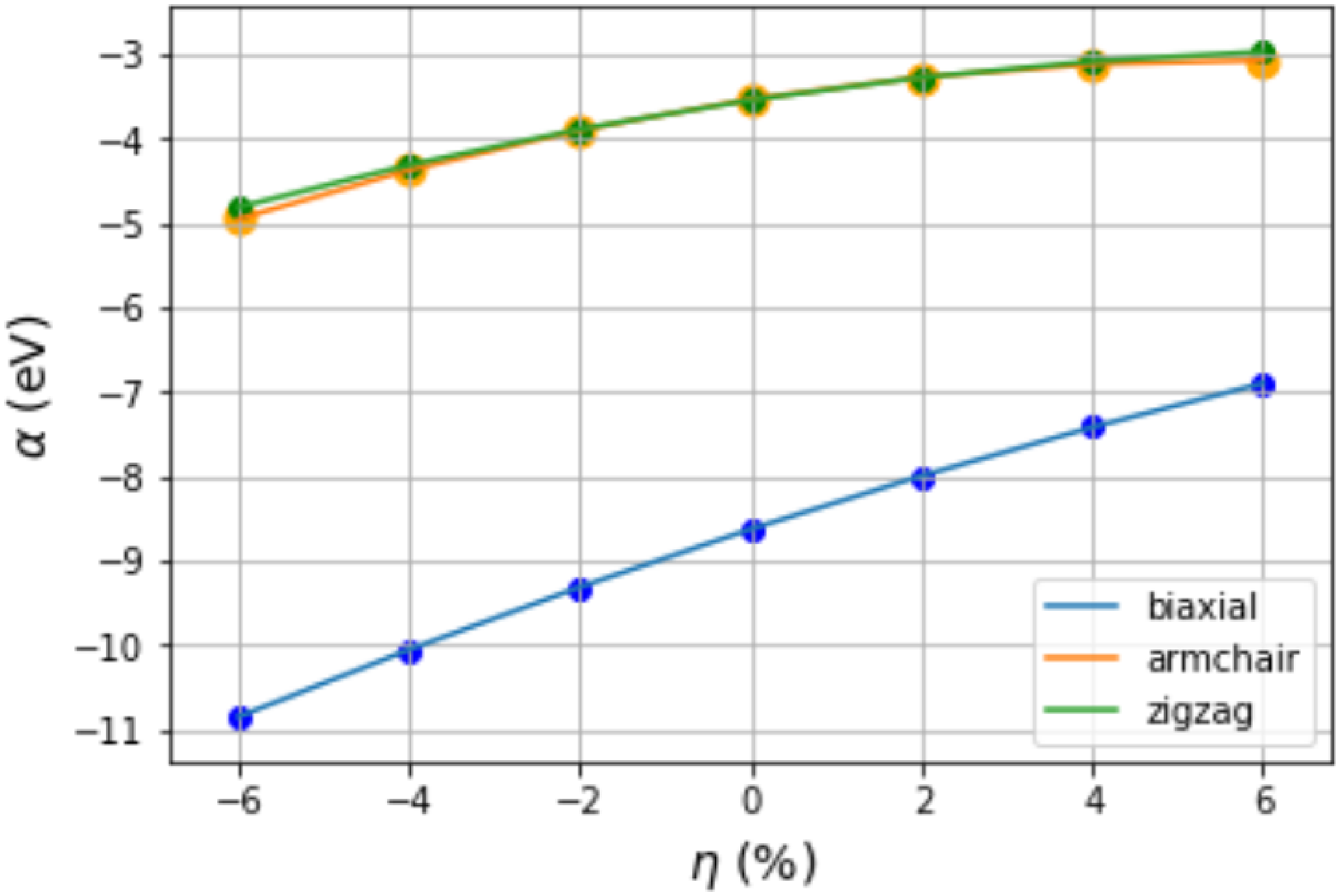}
 }
\caption{Results for graphene.
Top panel: Work function dependence on the biaxial (blue dots),
the uniaxial-zigzag (green dots), and the uniaxial-armchair (orange dots) strain
from $-16\%$ (compressive) to $+16\%$ (tensile).
Bottom panel: Deformation potential $\alpha$ associated with biaxial and uniaxial strains.}
\label{graphene_results}
\end{figure}

The WF dependence of graphene on biaxial and uniaxial strains  is computed for values of strain
in the range from $-16\%$ to  $+16\%$.
For the uniaxial cases, deformation in both zigzag and armchair directions is  investigated taking into account  the associated PR.
Results are reported in the top panel of Fig.~\ref{graphene_results} and show good agreement with the literature \cite{Giovannetti2008,Choi2010PRB,Ziegler2011PRB,Batrakov2019,Legesse2017,Yang2017,Postorino2020}.

We observe that the WF grows as the tensile strain increases.
A possible explanation for this trend is based on the fact that, as long as the material is stretched, the interaction among the ions of the lattice decreases.
In this way the system approaches the behavior of isolated atoms in the limit of infinite tensile deformation. The ionization potential for the C atom is \SI{11.2}{eV}, a value  much higher than the  WF  of  its  corresponding two  dimensional  form  at equilibrium,
\SI{4.25}{eV} in PBE (\SI{4.48}{eV} in LDA).
For this reason, it is expected that the WF of graphene, characterized  by  fully  covalent  bonds,  should  grow  with increasing uniform tensile deformation.

We also observe that the change rate of the WF depends on the type of strain applied. In particular, uniaxial strains highlight an almost identical behavior between the zigzag and armchair directions, with small differences
only for high values of the deformation. For biaxial strain a steeper change rate is observed.
This difference is due to the fact that, for a given value of $\eta$, the lattice deformation  is larger for biaxial strain since the atoms are uniformly displaced in all directions.

Knowing the WF dependence of the strain we  extract the deformation potential $\alpha$, defined as
\begin{equation}\label{alpha_def1}
\alpha_{s} = -\frac{dW_{s}^{\eta}}{d\eta}.
\end{equation}
Here, $W_s^{\eta}$ indicates the value of the WF corresponding to a strain of
magnitude $\eta$ and the index $s$ takes the values $\mathrm{bi,arm}$ and $\mathrm{zig}$.
We emphasize that for the uniaxial armchair and zigzag strains the deformation potential in  Eq.~(\ref{alpha_def1}) is
calculated taking into account the PR.

The {\it ab initio} estimate of Eq.~(\ref{alpha_def1}) is evaluated through a polynomial fitting of the WF
curves in the top panel of Fig.~\ref{graphene_results} and by the subsequent computation of the first derivative for each value of strain.
This procedure is restricted to a limited range (from $-6\%$ to $+6\%$), since the WF curves become less smooth for higher strain values, and the numerical extraction of the derivative becomes less trivial.
Corresponding results, reported in the bottom panel of Fig.~\ref{graphene_results}, show that in graphene the zigzag and armchair deformation potentials are practically identical, and much weaker than the biaxial one, for the reasons discussed above.

\subsection{Strain dependence of the WF and  extraction of deformation potential for silicene}
\label{sec:silicene}

\begin{figure}[!h]
\raggedleft{
\includegraphics[width=1.\linewidth]{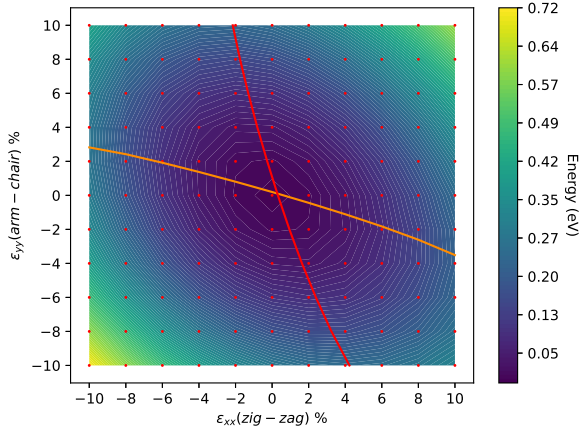}
}
\caption{Plot of the total energy of silicene against the values of $\varepsilon _{xx/yy}$.
        The dark orange (red) curve represents the value of the $\varepsilon_{yy}$
        ($\varepsilon_{xx}$) strain that minimizes the energy of the system given the constraint
        $\varepsilon _{xx} (\varepsilon _{yy}) = \mbox{const}$.
        The red dots represent  the sampled combinations of $\epsilon_{xx/yy}$ values used to compute  the total energy of the system.
        }
\label{fig:silicene_poisson}
\end{figure}

The same analysis described for graphene is performed also for silicene, using the PBE functional for both
tensile and compressive deformations. The optimized equilibrium lattice constant is found to be equal to
\SI{3.867}{ \AA} with a buckling of \SI{0.448}{\AA}. The WF for zero strain
is obtained to be \SI{4.35}{eV} (a test calculation in LDA gives a value of \SI{4.76}{ eV}). Also for silicene, the PR of the system is computed {\it ab
initio} by employing the same procedure used for graphene (see Fig.~\ref{fig:silicene_poisson}).

\begin{figure}[!h]
\raggedleft{
\includegraphics[width=1.\linewidth]{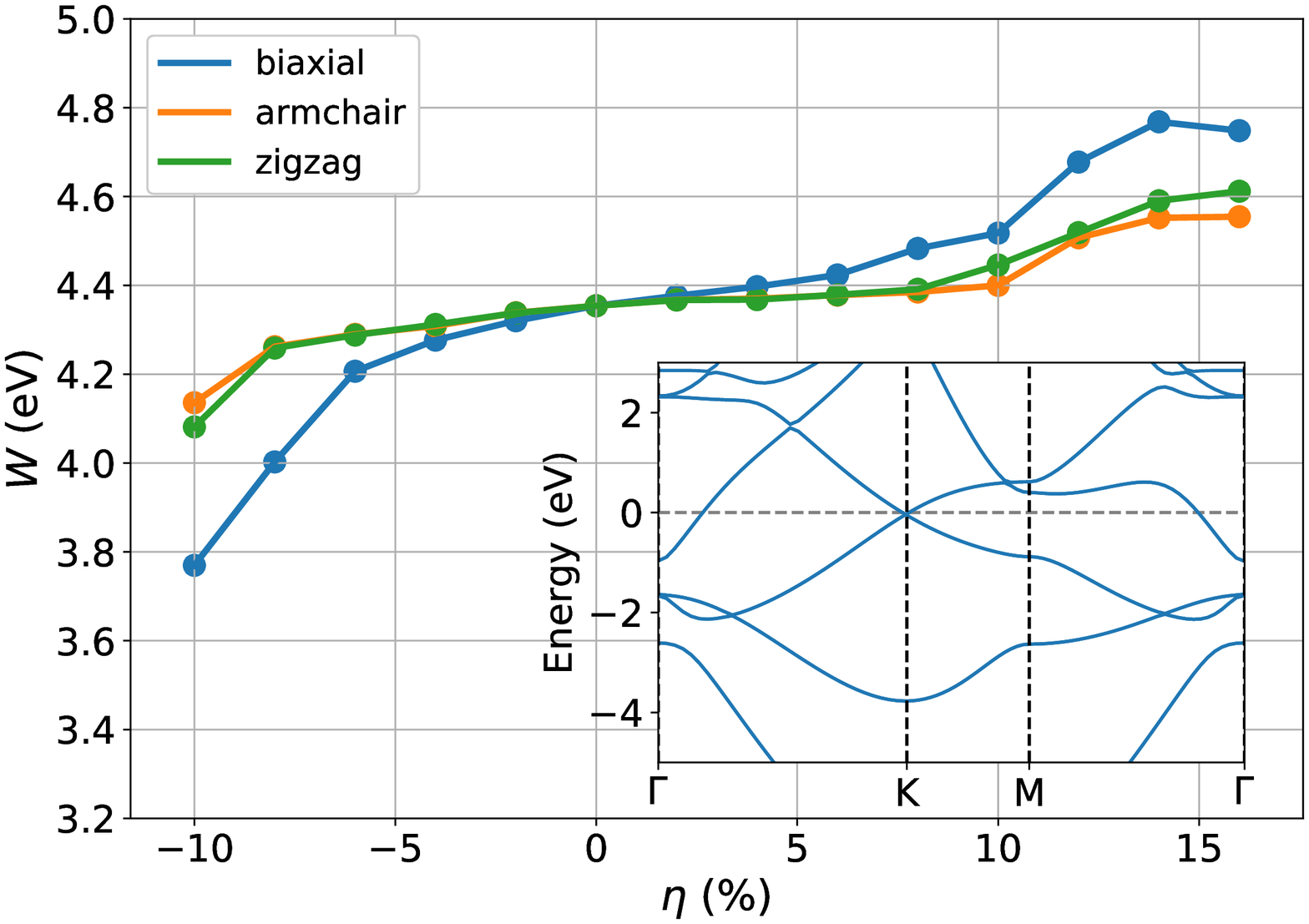}
\includegraphics[width=1.\linewidth]{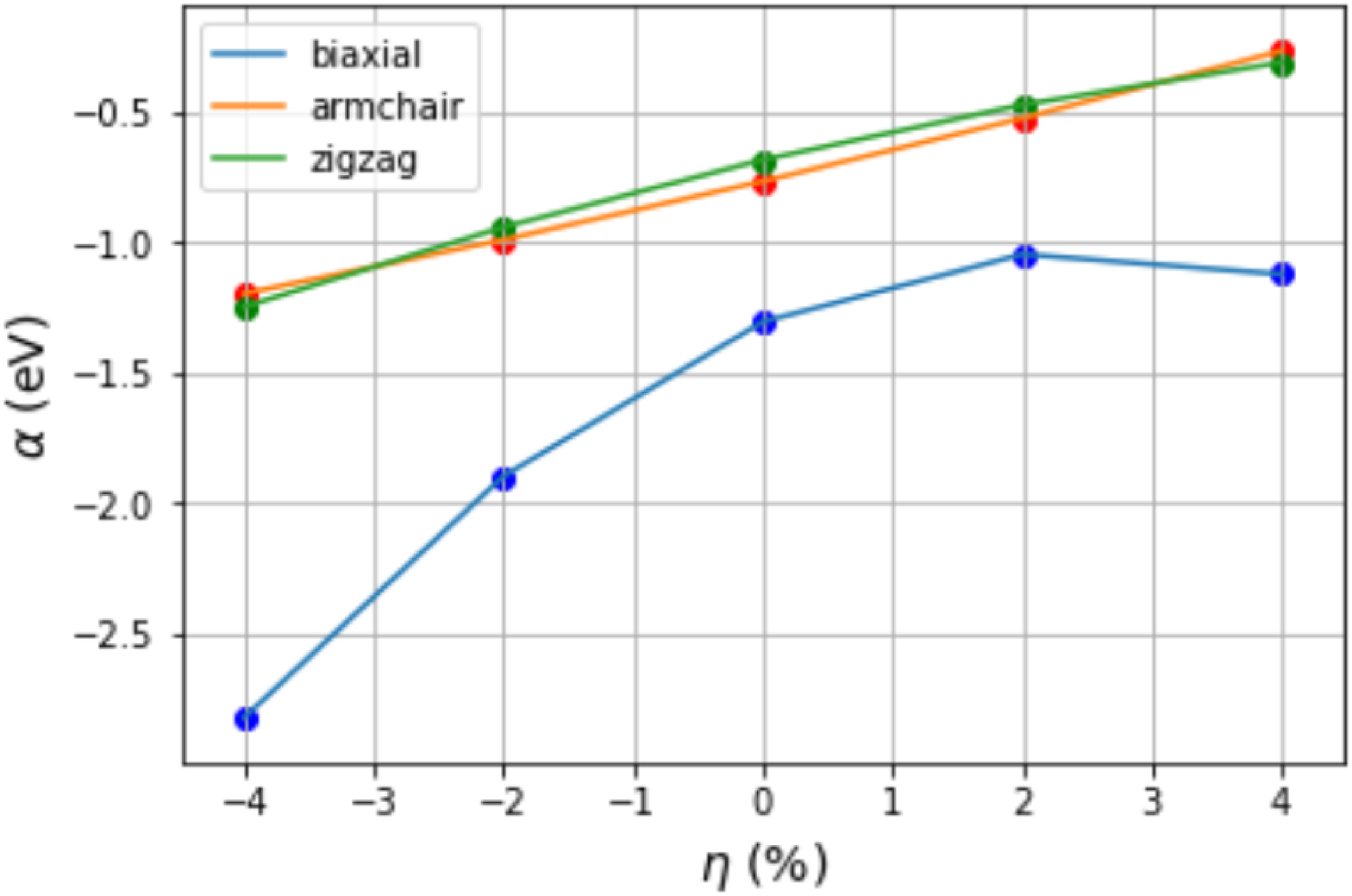}
}
\caption{Results for silicene.
Top panel: Work function dependence on the biaxial (blue dots),
the uniaxial-zigzag (green dots), and the uniaxial-armchair (orange dots) strain. The inset shows the bands structure of the system with a biaxial strain of $12\%$.
Bottom panel: Deformation potential $\alpha$ associated with biaxial and uniaxial strains. }
\label{silicene_result}
\end{figure}

Results are reported in Fig.~\ref{silicene_result} and show the strain dependence for both the WF and
the deformation potential. In particular, the behavior of the WF is presented for values of strain that range from
$+16\%$,  in the tensile region, down to $-10\%$, in the compressive part. The curves present a smooth behavior only for
values of strain limited from $-4\%$ up to  $+4\%$.
This non-monotonic behavior of the WF is due to the fact that, for high values of strain (compressive or tensile)
there is a change in the electronic band structure.
In particular, for high tensile strains, an empty band goes down in energy and crosses the Fermi level close to $\Gamma$
[see the inset in Fig.~\ref{silicene_result}].
On the other hand, for high compressive strains, a filled band near $\Gamma$ increases
its energy and crosses the Fermi level from below. All these major modifications
of the electronic band structure shift the Fermi level away from the Dirac point.
This causes a nonregular trend of the work function vs strain in the case of silicene. In graphene, instead,
this does not happen (at the considered strain values)
because the gap at $\Gamma$ is much larger than the silicene one.

The obtained data show that, analogously to what happens for graphene, the WF increases with increasing stretching. This increasing occurs since in the limit of infinite
tensile strain the WF tends towards the atomic ionization potential, \SI{8.1}{eV} in silicon.

We  calculate the deformation potential by extracting the numerical derivative of the WF in the range of  -4\% up to  +4\% .
A comparison with the results for graphene highlights that the values of the deformation potential obtained for silicene are much smaller in all the considered cases.

\section{Deformation potential in the tight-binding strained Hamiltonian}
\label{sec:deformation-potential}

In this section we relate the obtained above results to the parameters of the tight-binding
Hamiltonian. The electrons in the valence and conduction bands of graphene and silicene are described by the following Hamiltonian
\begin{equation}
\label{eqn:hamiltonian_host}
\mathrm{H}=\mathrm{H}_{\mathrm{hop}}+\mathrm{H}_{\mathrm{pot}}.
\end{equation}
Here $\mathrm{H}_{\mathrm{hop}}$ is the conventional tight-binding Hamiltonian  for $\pi$ orbitals
that describes hopping between nearest-neighbors \cite{Vozmediano2010PR,Amorim2016PR,Naumis2017RPP}.
We do not write down its explicit form since the corresponding hopping parameters
and their strain dependence are not considered in the present work.

The main interest for us represents the potential term
\begin{equation} \label{eqn:hamiltonian_pot}
\mathrm{H}_{\mathrm{pot}}   =
\sum_{i, \delta} U^{\phantom\dagger}_{\delta i}
\hat{c}^\dagger_{\delta i} \hat{c}^{\phantom\dagger}_{\delta i},
\end{equation}
where  $i$ run over $N/2$ lattice cells, indices  $\delta=A$ and $B$ enumerate the sublattices,
operator $\hat{c}_{\delta,i}^\dagger$ ($\hat{c}_{\delta, i}$)
creates (annihilates) an electron at the corresponding lattice site,
the spin index is omitted for brevity,
and $U_{\delta i}$ is the on-site deformation-dependent potential. $U_{\delta i}$ consists of the strain-independent part $E_\mathrm{D}$, which determines the energy of the Dirac point in unstrained graphene, and the strain-dependent part.

For uniform strain, the potential energy does not depend on the lattice site.
Assuming also the linear dependence of the on-site energy on strain, one can rewrite
the Hamiltonian (\ref{eqn:hamiltonian_pot}) as follows:
\begin{equation}
\label{H_pot}
\mathrm{H}_{\mathrm{pot}} = \sum_{i, \delta} (\alpha_{xx} \varepsilon_{xx}
+  \alpha_{yy} \varepsilon_{yy}+ E_\mathrm{D} )
c^\dagger_{\delta i} c^{}_{\delta i}.
\end{equation}
Here we introduced two deformation potential constants: $\alpha_{xx}$ and $\alpha_{yy}$.
The values of these constants are determined from the strain dependence of the WF:
\begin{equation}
\alpha_{xx} = - \left(\frac{\partial W}{\partial \varepsilon_{xx} } \right)_{\varepsilon_{yy} =0}, \quad
\alpha_{yy} = - \left( \frac{\partial W}{\partial \varepsilon_{yy} } \right)_{\varepsilon_{xx} =0},
\end{equation}
where $xx$ and $yy$ correspond to the armchair and zigzag directions, respectively.
The range of validity of Eq.~(\ref{H_pot}) follows from the results presented in
Figs.~\ref{graphene_results} and \ref{silicene_result} and discussed in Sec.~\ref{sec:ab-results}. For graphene this corresponds to strain values from $-6\%$ to $6 \%$, and for silicene  this corresponds to strain values from $-4\%$ to $4 \%$.

In Sec.~\ref{sec:ab-results} we determined the $\alpha_{s}$ values,
which describe the deformation of the samples in the presence of
Poisson's transverse contraction characterized by  $\nu_{s}$ with $s = \mathrm{arm}$ and $\mathrm{zig}$.
These parameters can be related to each other by taking into account that
\begin{equation}
\label{system}
\begin{split}
\frac{d W^\eta_{\mathrm{arm}}} {d \eta} & = \frac{\partial W}{\partial \varepsilon_{xx} }-
\nu_{\mathrm{arm}} \frac{\partial W}{\partial \varepsilon_{yy} } , \\
\frac{d W^\eta_{\mathrm{zig}}} {d \eta} & = - \nu_{\mathrm{zig}} \frac{\partial W}{\partial \varepsilon_{xx} }+
\frac{\partial W}{\partial \varepsilon_{yy} } .
\end{split}
\end{equation}
Now we assume that the corresponding derivatives are constants for small values of the strain.
Solving the system (\ref{system}) one obtains
\begin{equation}
\begin{split}
\alpha_{xx} & = \frac{\alpha_{\mathrm{arm}} + \nu_{\mathrm{arm}} \alpha_{\mathrm{zig}} }
{1 - \nu_{\mathrm{arm}} \nu_{\mathrm{zig}}}, \\
\alpha_{yy} & = \frac{\alpha_{\mathrm{zig}} + \nu_{\mathrm{zig}} \alpha_{\mathrm{arm}}}
{1 - \nu_{\mathrm{arm}} \nu_{\mathrm{zig}}}.
\end{split}
\end{equation}
The values of the constants $\alpha_{s}$, $\alpha_{xx,yy}$ and the PR for the tensile strain
for graphene and silicene are provided in Table~\ref{table1}.

\begin{table}
    \begin{tabular}{|c||c|c|c|  }
        \hline
        \! \! Graphene \!\!  & \!\! Armchair \!\! & \!\! Zigzag \!\! & \!\! Biaxial \!\! \\
        \hline
        $\alpha_{s} $ (eV)   & -3.5  & -3.5 &  -8.6\\
        \hline
        $\nu_{\mathrm{arm,zig}} $    & 0.14  & 0.14 &  --\\
        \hline
        $\alpha_{xx,yy} $ (eV) for $\nu=0$   & -4.1  & -4.1 &  --\\
        \hline \hline
        \!\! Silicene \!\!  & Armchair & Zigzag & Biaxial \\
        \hline
        $\alpha_{s} $ (eV)  & -0.8  & -0.7 &  -1.3\\
        \hline
        $\nu_{\mathrm{arm,zig}} $    & 0.22  & 0.13 &  --\\
        \hline
        $\alpha_{xx,yy} $ (eV) for $\nu=0$   & -0.71  & -0.49 &  --\\
        \hline
    \end{tabular}
    \caption{
        The values of deformation potential constants  $\alpha_{s}$, with $s = \mathrm{bi,arm,zig}$;
        $\alpha_{xx,yy}$; and the Poisson's ratios $\nu_{arm,zig}$ (for the tensile strain)
        for graphene and silicene, calculated in the vicinity of $\eta =0$.
        }\label{table1}
\end{table}

\section{Collapse of Landau levels}
\label{sec:discussion}

As  mentioned in the Introduction, the energy spectrum of
the graphene Dirac fermions  in crossed external magnetic and electric fields \cite{Lukose2007PRL,Peres2007JPCM}
is given by Eq.~(\ref{LL}), with the corresponding Landau scale shown by Eq.~(\ref{Landau-scale}).
It is easy to see from these equations that, in the case of a constant value of the
in-plane electric field $E$,  the Landau levels would collapse
as the magnetic field reaches the value $B_c = E/v_F $ from above.
Some indications of this effect have been obtained experimentally in Refs.~\cite{Vibnor2009PRB,Gu2009PRL}.

Interestingly,  in Dirac materials the strain can induce the same phenomena.
The experimental observation of the Landau levels induced by inhomogeneous strain \cite{Levy2010Science} is probably the most spectacular effect associated with straintronics  \cite{Levy2010Science}.
The key point is that the strain-induced change in the hopping energy between neighboring atoms in the Hamiltonian $\mathrm{H}_{\mathrm{hop}}$
can be described by some kind of vector potential $\mathbf{A}_{\mathrm{pm}}$
(see Refs.~\cite{Vozmediano2010PR,Amorim2016PR} for a review). For the $x$-axis aligned in the armchair direction \cite{Kitt2012PRB},
it reads
\begin{equation}
\label{pseudo-vector}
\mathbf{A}_{\mathrm{pm}} = \frac{\hbar \beta }{2 a_0}  \left(
               \begin{array}{c}
                 2 \varepsilon_{xy} \ \\
                 \varepsilon_{xx} - \varepsilon_{yy} \\
               \end{array}
             \right),
\end{equation}
where $\beta$ is the dimensionless Gr\"{u}neisen parameter for the lattice deformation
and $a_0$ is the lattice constant.
The generic position-dependent  strain tensor $\varepsilon_{i j}$, with $i,j = x,y$, is
related to the displacement  (\ref{eq:strain2}) by the relation $\varepsilon_{ij} = (\partial_i u_j + \partial_j u_i) /2$.

The  vector potential, Eq.~(\ref{pseudo-vector}), generates a pseudomagnetic field
$\mathbf{B}_{\mathrm{pm}} = \mathbf{\nabla} \times \mathbf{A}_{\mathrm{pm}}$.
It formally resembles a real magnetic field, with the crucial distinction that
it is directed oppositely in $\mathbf{K}$ and $\mathbf{K}^\prime$  valleys.
The sign of the  pseudomagnetic field depends on the  valley, and, for example,
in the $\mathbf{K}$ valley,
\begin{equation}
\label{PMF}
B_{\mathrm{pm}} = \frac{\hbar \beta  }{a_0} \left( \frac{1}{2} \partial_x (\varepsilon_{xx} - \varepsilon_{yy})  -
\partial_y \varepsilon_{xy} \right),
\end{equation}
whereas it has the opposite sign in the $\mathbf{K}^\prime$ valley.

Then, the deformation potential part of the Hamiltonian $\mathrm{H}_{\mathrm{pot}}$, Eq.~(\ref{H_pot}), contains
the scalar potential $\alpha A_0 = \alpha (\varepsilon_{xx} + \varepsilon_{yy})$,
which has the same sign in
both the $\mathbf{K}$ and $\mathbf{K}^\prime$  valleys. Accordingly, the deformation potential acts as an
electric field per unit charge
$ E_i = - \alpha_{ik} \partial_k A_0$.
Bearing in mind the isotropic graphene (see Table~\ref{table1}),
we  assume that  $\alpha_{xx} = \alpha_{yy} = \alpha$.
Yet, since in silicene $\alpha_{xx} \neq \alpha_{yy}$,
the results presented below are not directly applicable.

One can see that  uniform strain results in the appearance of a constant strain-induced vector potential corresponding to
the shift of the  $\mathbf{K}$ and $\mathbf{K}^\prime$ points, so that the pseudomagnetic field is zero.   Since $A_0$ is position independent, also
an electric field is  absent.

On the other hand, creation of the pseudo Landau levels requires a special configuration with inhomogeneous strain.
To simplify  theoretical modeling, the pseudo Landau levels are very often treated assuming that  the deformation is a pure shear,
so that $A_0 = \varepsilon_{xx} + \varepsilon_{yy} =0$, and the corresponding term in the Hamiltonian does not appear.
This assumption is rather unphysical, and when the deformation potential is included, new effects are expected. For some strain configurations,
the deformation potential acts as an in-plane electric field.

A special strain configuration was considered in Ref.~\cite{Castro2017PRB}. In our notations  it can be written
as  $\varepsilon_{xx} = 2 a_0 B x/(\beta \hbar)$, $\varepsilon_{yy} = 0$, and $\varepsilon_{xy} =0$.
It corresponds to  the strain-induced vector and scalar potentials
$A_{\mathrm{pm}} = (0,B x)$ and $A_0 = 2 a_0 B x/(\beta \hbar)$, respectively.
Evidently, they generate crossed
constant pseudomagnetic and electric fields of the magnitudes $B$ and $E_x = - 2 a_0 \alpha B/(\beta \hbar)$.
Then, the condition of the Landau level collapse acquires the form
$|\alpha| \gtrsim v_F \beta \hbar/(2a_0)$  \cite{Lukose2007PRL,Castro2017PRB}.
One can see that in this case the condition for the collapse
depends on the material constants $\alpha$, $\beta$, and $v_F$, which
 cannot be tuned easily.

Here we propose a different experimental setup with a special strain configuration that generates only the electric field, while
the pseudomagnetic field is absent. Then, applying a real magnetic field, one should be able to realize the Landau level collapse.
In fact, we obtain a pseudomagnetic  field of $B_{\mathrm{pm}} =0$ when in Eq.~(\ref{pseudo-vector}) the components of the pseudo vector potential are constants, i.e.,
$A_x = C_1$ and $A_x = C_2$. Then it is easy to see that this is possible when the components of the displacement
vector $u_{x,y} (x,y)$ satisfy the two-dimensional Laplace equations:
\begin{equation}
\label{Laplace}
\begin{split}
\frac{\partial^2 u_x}{\partial x^2} + \frac{\partial^2 u_x}{\partial y^2} & = 0,\\
\frac{\partial^2 u_y}{\partial x^2} + \frac{\partial^2 u_y}{\partial y^2} & = 0.
\end{split}
\end{equation}
Any harmonic function satisfies Eq.~(\ref{Laplace}), so one can consider the simplest nontrivial example:
\begin{equation}
\label{Laplace-solution}
\begin{split}
u_{x}(x,y) & =  d (x^2 - y^2) + h_1 x + h_2 y,\\
u_{y}(x,y) & =  2 d xy  + h_3 x + h_4 y.
\end{split}
\end{equation}
Here $d$ is a constant that has the dimension of an inverse length, while $h_{1,2,3,4}$ are the dimensionless constants
that describe the uniform strain.
\begin{figure}[!h]
\raggedleft{
\includegraphics[width=1.\linewidth]{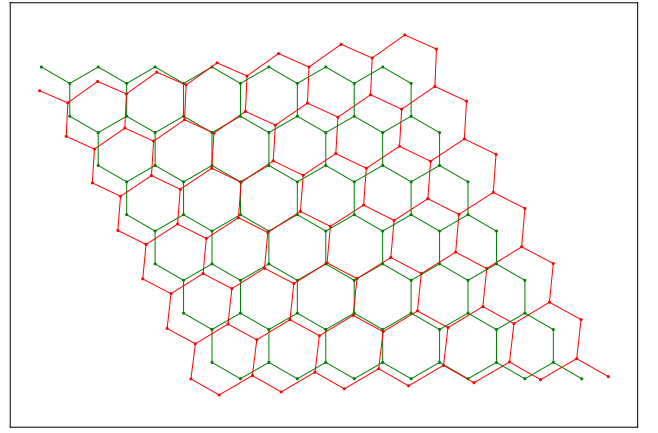}}
\caption{Example of a nonuniform strained lattice (red) built applying   Eq.~(\ref{Laplace-solution}) to ideal graphene (green). We  used $d a_0=0.01$ and $h_{1,2,3,4}=0.1$.}
\label{fig:strained_geo}
\end{figure}
This strain configuration, shown in Fig.~\ref{fig:strained_geo}, generates the following potentials:
\begin{equation}
\begin{split}
A_x = & \frac{\beta}{2 a_0} (h_2 + h_3), \qquad  A_y =    \frac{\beta}{2 a_0} (h_1 - h_4), \\
& A_0 = 4 d x + h_1 + h_4.
\end{split}
\end{equation}
One can see that this potential corresponds  to $B_{\mathrm{pm}} =0$  and a constant electric field
$E_x = -4 \alpha d /e$, where we explicitly included the electric charge $e$.
The constant  term $\alpha (h_1 + h_4)$ in $\mathrm{H}_{\mathrm{pot}} $ corresponds
to the uniform strain considered in the previous sections.

\begin{figure}[!h]
\raggedleft{
\includegraphics[width=1.\linewidth]{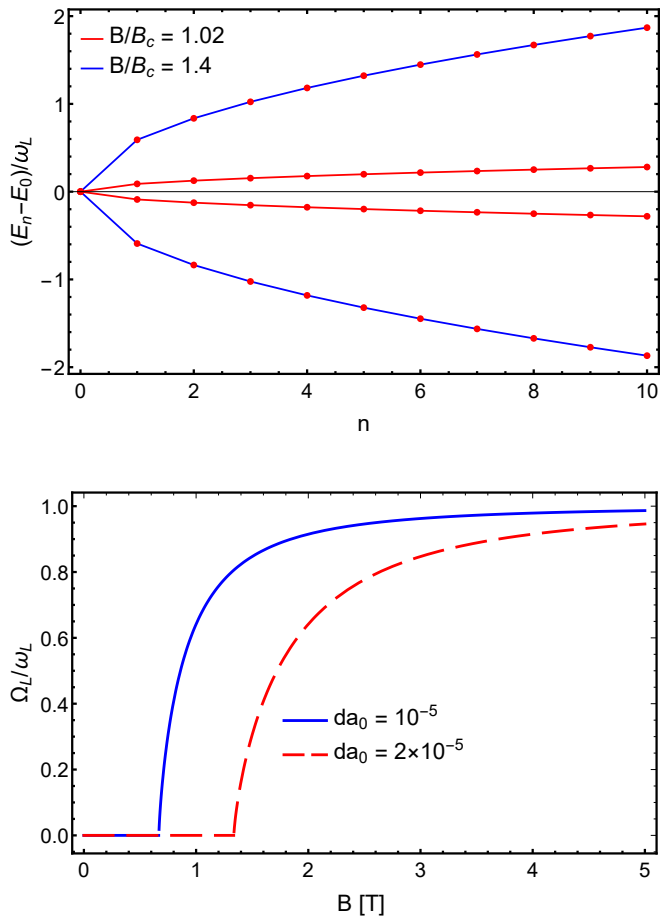}}
\caption{Top panel: the Landau level energies $(E_n - E_0)/\omega_L$
for $B/B_c = 1.4$ and $1.02$. Bottom panel: the dependence of the Landau scale
$\Omega_L/\omega_l$ on the magnetic field $B$ for $d a_0 = 10^{-5}$ and $2 \times 10^{-5}$.}
\label{fig:LL-collapse}
\end{figure}

When a constant external magnetic field is applied in addition to the strain induced electric field,  the condition of the Landau levels collapse $E = v_F B_c$ \cite{Lukose2007PRL} acquires the following form
\begin{equation}
\label{critical}
B_c = \frac{4 (d a_0)  |\alpha| }{e a_0 v_F} .
\end{equation}
Thus, as the magnetic field decreases to this critical value $B_c$,  the collapse occurs.
This is  illustrated in the top panel of Fig.~\ref{fig:LL-collapse}, where the energies of strain-affected Landau levels
$E_n - E_0$ in the units of the  Landau scale $\omega_L$
are shown  for the two values of $B/B_c$. They tend to
the energy of the lowest Landau level as $B$ approaches $B_c$.

Taking into account that $\SI{1}{T} = \SI{}{V\cdot s/m^2}$ and using the value of the Fermi velocity $v_F = \SI{1E6}{m/s}$ and
$a_0 = \SI{2.46}{~\AA}$, one obtains
the following estimate:
\begin{equation}
 B_c =  \SI{1.63E4} (d a_0)  |\alpha| ,
\end{equation}
where  $B_c$ is expressed in Tesla, while $\alpha $ is measured in eV.

Assuming a value of $d a_0 = 10^{-5}$, one finds that the Landau levels collapse  for graphene (see Table~\ref{table1}) would occur at $B_c = \SI{0.67}{T}$. Although for silicene the value of the critical field  is expected to be
dependent on the direction of the electrical field, one can make a rough estimate of $B_c$ using  Eq.~(\ref{critical}). Assuming
a value ща $|\alpha|=0.6$ eV given by the average of $\alpha_{xx}$ and $\alpha_{yy}$, it  gives the critical field $B_c= \SI{0.19}{T}$. Here we took $v_F = \SI{0.5E6}{m/s}$ and the same parameter $d$ as used for graphene.

The dependence of the Landau scale $\Omega_L$ of graphene,
for  two values of the dimensionless parameter $d a_0$ is presented in the bottom panel of Fig.~\ref{fig:LL-collapse}. This example shows that the measurements of the critical field $B_c$ can be used to extract the value of $d a_0$ characterizing the specific
of the deformation.
The estimates of $B_c$ confirm that the experiment, where the electric field is generated by the
non-uniform strain and an external real magnetic field $B$ is tuned to its critical value, can be implemented in practice.

\section{Conclusions}
\label{sec:concl}

In this work we show by {\it ab initio} calculations how the WFs of graphene and silicene depends on uniform compressive
and tensile strains. For small deformations the dependence is linear and corresponding values of the deformation potential
parameters are provided in  Table~\ref{table1}.
In accordance with both the experiment \cite{He2015APL} and  the {\it ab initio} results  the WFs of graphene and silicene
increase under the tensile strain. For small values of strain the armchair and zigzag deformation potentials turn out to be practically identical and approximately correspond to one half of the deformation potential associated with biaxial strain.

It has to be noted that strain tuning of the WFs of different materials has been a topic of research for a long time.
As an example we refer to the experiment  In Ref.~\cite{Li2004PM}
that shows the opposite strain dependence of the WFs in Cu and Al, viz., in the elastic range, tensile strain
results in the decrease of the WF. The corresponding  {\it ab initio} calculation that agreed with the
experiment was presented in Ref.~\cite{Pogosov2008}. Thus one of the questions for the future is to address how the corresponding  strain dependence of the WF is material dependent.

Finally,  we propose the experimental setup with a special strain configuration that generates only an electric field,
whereas the pseudomagnetic field is absent. In this case, in order to obtain the Landau level staircase,  an external magnetic
field should be applied.
Such a setup allows one to explore the phenomenon of the Landau levels collapse more easily, since the strain-induced electric field
and the magnetic field can be controlled independently.

\section*{acknowledgments}

We acknowledge the support of EC for the
HORIZON 2020 RISE “CoExAN” Project (Project No. GA644076).
V.P.G. and S.G.Sh. acknowledge a partial support
by the National Academy of Sciences of Ukraine grant  ``Functional Properties of Materials Prospective for Nanotechnologies'' (project No. 0120U100858). They are also grateful to V.M.~Loktev and Y.V.~Skrypnyk for illuminating discussions. D.G. and O.P. acknowledge the
EC for support through the HORIZON 2020 RISE  "DiSeTCom" project (GA823728).

\bibliography{references}

\end{document}